\documentclass[3p]{elsarticle}
\usepackage{amsmath,amsfonts,amssymb,graphicx,color}
\usepackage{subfigure}

\begin{document}
\begin{frontmatter}

\title{Ultra High Energy Cosmic Rays: The disappointing model}

\author[LNGS]{R. Aloisio}
\author[LNGS,CFA]{V. Berezinsky}
\author[CFA,IP]{A. Gazizov\corref{cor1}}
\ead{askhat.gazizov@lngs.infn.it}
\cortext[cor1]{Corresponding author}

\address[LNGS]{INFN, National Gran Sasso Laboratory, I-67010 Assergi (AQ), Italy}
\address[CFA]{Gran Sasso Astroparticle Center, I-67010 Assergi (AQ), Italy}
\address[IP]{B.I. Stepanov Institute of Physics of NASB, 68 Independence Avenue,
BY-22072 Minsk, Belarus}

\begin{abstract}
We develop a model for explaining the data of Pierre Auger Observatory
(Auger) for Ultra High Energy Cosmic Rays (UHECR), in particular, 
the mass composition being steadily heavier with increasing energy from 
$3$~EeV to $35$~EeV. The model is based on the proton-dominated 
composition in the energy range (1 - 3)~EeV observed in both Auger and 
HiRes experiments. Assuming extragalactic origin of this component, we 
argue that it must disappear at higher energies due to a low maximum 
energy of acceleration, $E_p^{\max} \sim (4 - 10)$~EeV. Under an assumption
of rigidity acceleration mechanism, the maximum acceleration energy for a 
nucleus with the charge number $Z$ is $Z E_p^{\max}$, and the highest 
energy in the spectrum, reached by Iron, does not exceed ($100 - 200$)~EeV. 
The growth of atomic weight with energy, observed in Auger, is provided 
by the rigidity mechanism of acceleration, since at each energy 
$E=Z E_p^{\max}$ the contribution of nuclei with $Z' < Z$ vanishes. 
The described model has disappointing consequences for future observations 
in UHECR: Since average energies per nucleon for all nuclei are less than 
($2 - 4$)~EeV, (i) pion photo-production on CMB photons in extragalactic 
space is absent; (ii) GZK cutoff in the spectrum does not exist; 
(iii) cosmogenic neutrinos produced on CMBR are absent; (iv) fluxes of 
cosmogenic neutrinos produced on infrared - optical background radiation  
are too low for registration by existing detectors and projects. Due to 
nuclei deflection in galactic magnetic fields, the correlation with nearby 
sources is absent even at highest energies. 
\end{abstract}

\end{frontmatter}

\section{Introduction}%
\label{sec:introduction}%
There is a dramatic conflict between recent observational data of 
two largest UHECR detectors: HiRes \cite{Abbasi:2007sv} and Auger 
\cite{Abraham:2008ru}. The HiRes data confirm well signatures of 
{\em proton} propagation through cosmic microwave background radiation 
(CMBR), the GZK cutoff \cite{Greisen:1966jv,Zatsepin:1966jv} and the 
pair-production dip \cite{Berezinsky:1988wi,Berezinsky:2002nc,
Aloisio:2006wv,Berezinsky:2002vt,Berezinsky:2005cq}, together with a 
proton-dominated mass composition \cite{sokol-trondheim,Sokolsky:2010kb}. 
The Auger data strongly favor the nuclei composition getting progressively 
heavier in an energy range ($4 - 40$)~EeV and indicate a strong 
spectrum steepening at highest energies not much consistent with 
the predicted shape of the GZK cutoff.

Preliminary results of the new Telescope Array experiment 
\cite{TA,Thomson:2010tc} are in good agreement with the HiRes data both 
in spectrum, chemical composition and in the observed isotropy of UHECR 
arrival directions. 

\begin{figure*}[t]
\centering
\mbox{\subfigure{\includegraphics[height=50mm,width=70mm]{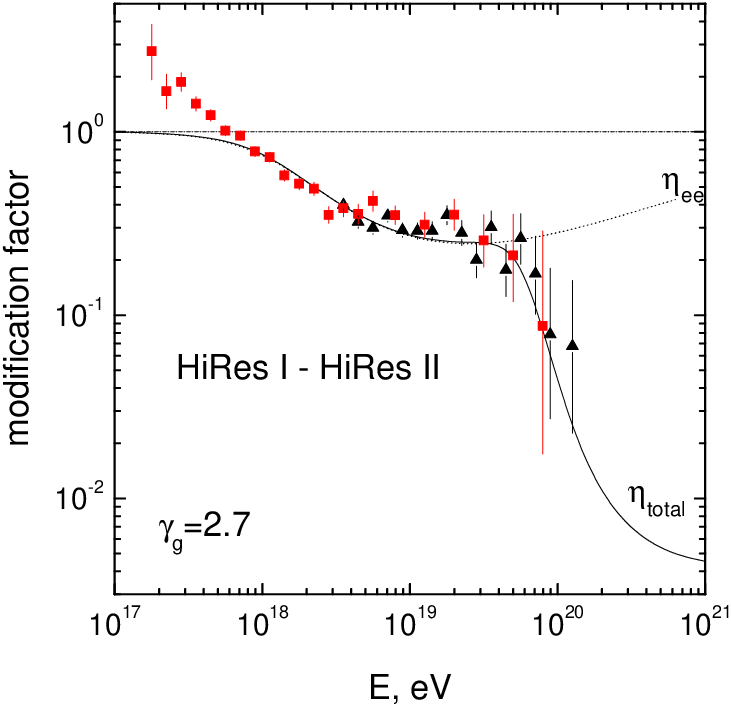}}\quad
\subfigure{\includegraphics[height=50mm,width=65mm]{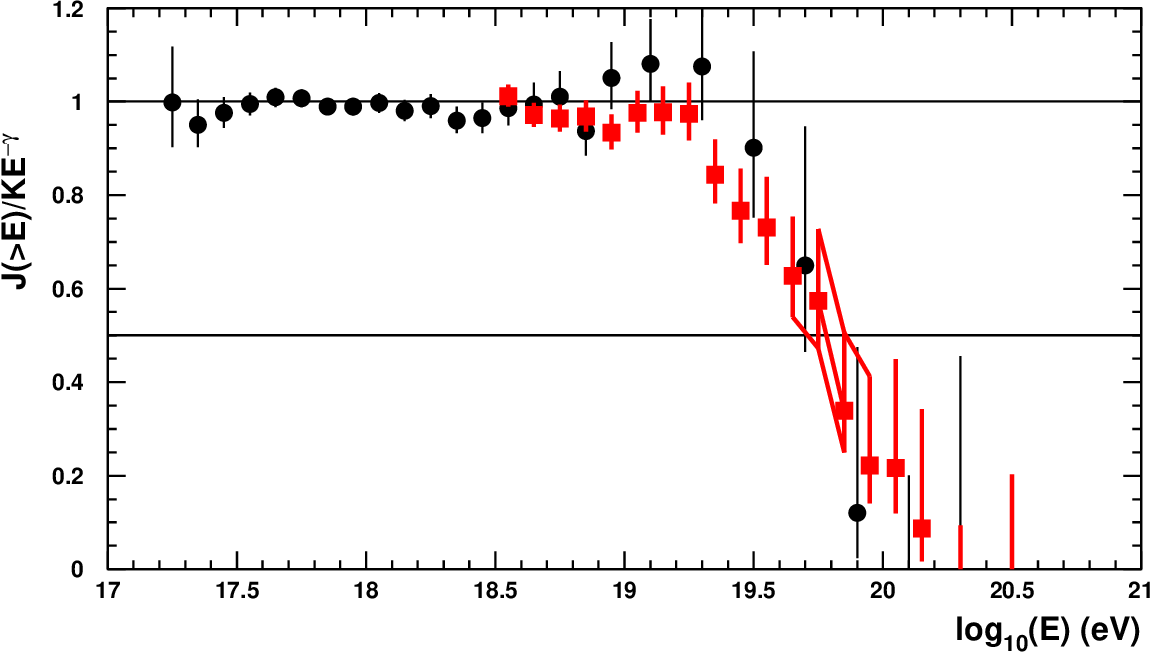}}}
\caption{
{\em Left panel}: Pair-production dip and GZK cutoff in 
terms of modification factor in comparison with the HiRes observational 
data \cite{Abbasi:2007sv} (HiRes 2 monocular -- boxes, HiRes 1 monocular 
-- triangles). Curves $\eta_{\rm tot}$ and $\eta_{ee}$ show 
the total spectrum and the spectrum calculated with only adiabatic and 
pair-production energy losses included, respectively. {\em Right panel}: 
$E_{1/2}$ as a numerical characteristic of the GZK cutoff in the 
integral HiRes spectrum (see text); HiRes 2 monocular -- circles , HiRes 
1 monocular -- boxes.
} 
\label{Fig1}%
\end{figure*}

We shall discuss first the HiRes data. In the left panel of Fig.~\ref{Fig1} 
the comparison of the HiRes data with the calculated pair-production dip 
and the GZK cutoff is shown in terms of modification factor $\eta (E)$. 
This quantity is given by a ratio of the energy spectrum $J_p(E)$ calculated 
with all energy losses taken into account, and the unmodified spectrum 
$J_p^{\rm unm}$, where only adiabatic energy loss (the red-shift) is included: 
$\eta(E)=J_p(E)/J_p^{\rm unm}(E)$. The modification factor is a in
convenient quantity, describing the dip and the beginning of the GZK cutoff 
practically model-independent way (see \cite{Berezinsky:2002nc,
Aloisio:2006wv}). These two spectral features are the signatures of protons 
interacting with CMBR. In the left panel of Fig.~\ref{Fig1}
one can see the good agreement of the HiRes spectrum with both dip and the
GZK cutoff. The nature of the spectrum steepening seen in Fig.~\ref{Fig1} 
as the GZK cutoff is further confirmed by the right panel of 
Fig.~\ref{Fig1} valid for the integral spectrum. 

In the integral spectrum the GZK cutoff is characterized by the energy 
$E_{1/2}$, where the calculated spectrum $J(>E)$ becomes half of the 
power-law extrapolation spectrum $K E^{-\gamma}$ from low energies. 
As calculations \cite{Berezinsky:1988wi} show, this energy is $E_{1/2} = 
10^{19.72}$~eV for a wide range of generation indices from 2.1 to 
2.8. HiRes collaboration found $E_{1/2} = 10^{19.73 \pm 0.07}$~eV in 
a good agreement with the theoretical prediction. In the right panel 
of Fig.~\ref{Fig1} we reproduce the HiRes graph \cite{Abbasi:2007sv} 
from which $E_{1/2}$ was determined. The plotted value 
is given by ratio of the measured flux $J(>E)$ and its power-law 
approximation $KE^{-\gamma}$. An extrapolation of this ratio to 
higher energies is given by 1 (unity), while the intersection of the 
measured ratio with the horizontal line $1/2$ gives $E_{1/2}$. 

With some caution one may conclude that HiRes has detected the 
signatures of proton interaction with CMBR in the form of 
pair-production dip and GZK cutoff. The final proof of this 
conclusion must come from the direct measurement of the mass 
composition of primaries. 

\begin{figure*}[ht]
\centering
\mbox{\subfigure{\includegraphics[height=50mm,width=65mm]{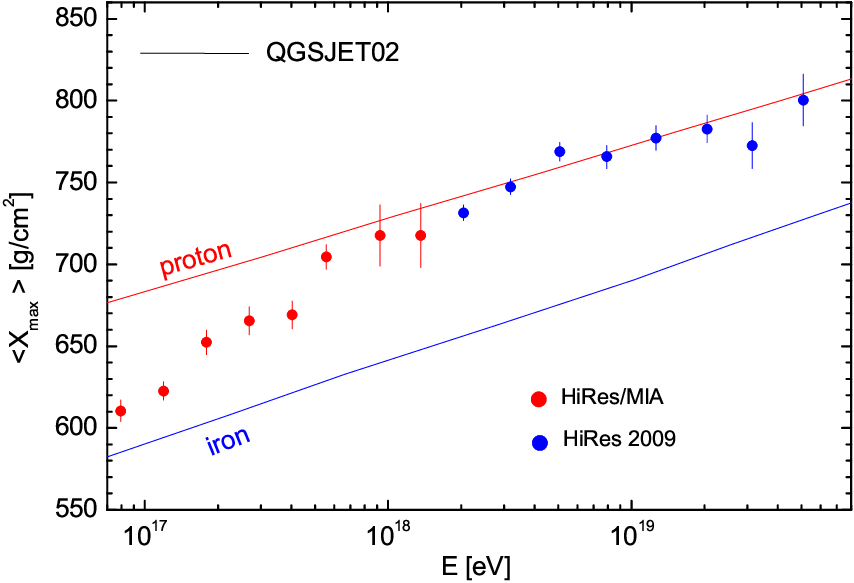}}\quad
\subfigure{\includegraphics[height=51mm]{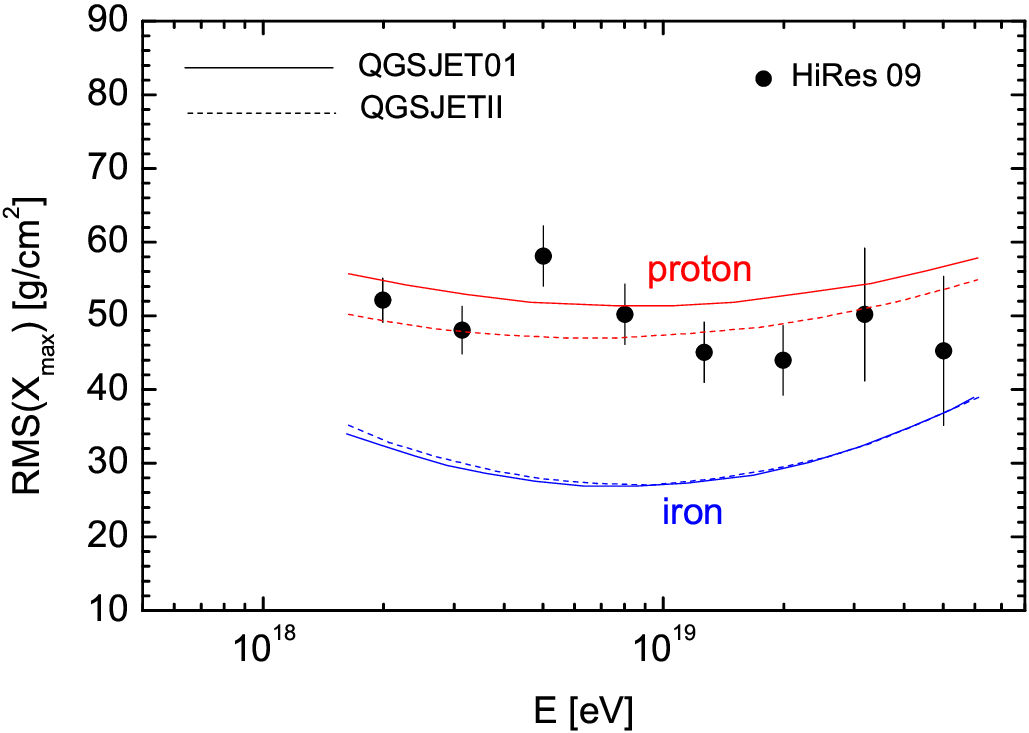}}}
\caption{
$X_{\rm max}$ as function of energy ({\em left panel}) and RMS($X_{\rm max}$), 
the width of distribution over $X_{\rm max}$, ({\em right panel}) from HiRes 
data \cite{sokol-trondheim, Sokolsky:2010kb}. The calculated values for 
protons and Iron are given according to QGSJETII model \cite{Ostapchenko:2005nj}. 
Note that the theoretical curves for RMS in the right panel have
different shape than in the right panel of Fig.~\ref{Fig3}. It is
caused by using in the HiRes analysis the truncated Gaussian
distribution over $X_{\max}$ \cite{sokol-trondheim,Sokolsky:2009Blois,Abbasi:2009nf}, 
while the Auger collaboration does not introduce any cuts.
}
\label{Fig2}%
\end{figure*}

The HiRes measurements of elongation rate and distribution 
over $X_{\rm max}$, the atmospheric height of the shower maximum, confirm 
the dominance of proton composition, indeed. In Fig.~\ref{Fig2} we 
plot the data of HiRes \cite{sokol-trondheim,Sokolsky:2010kb} on $X_{\rm max}$ 
as a function of energy (elongation curve) and RMS$(X_{\rm max})$, the width of 
distribution over $X_{\rm max}$. One can see that both quantities 
agree with proton-dominated composition.

\begin{figure*}[ht]
\centering
\mbox{\subfigure{\includegraphics[height=50mm,width=65mm]{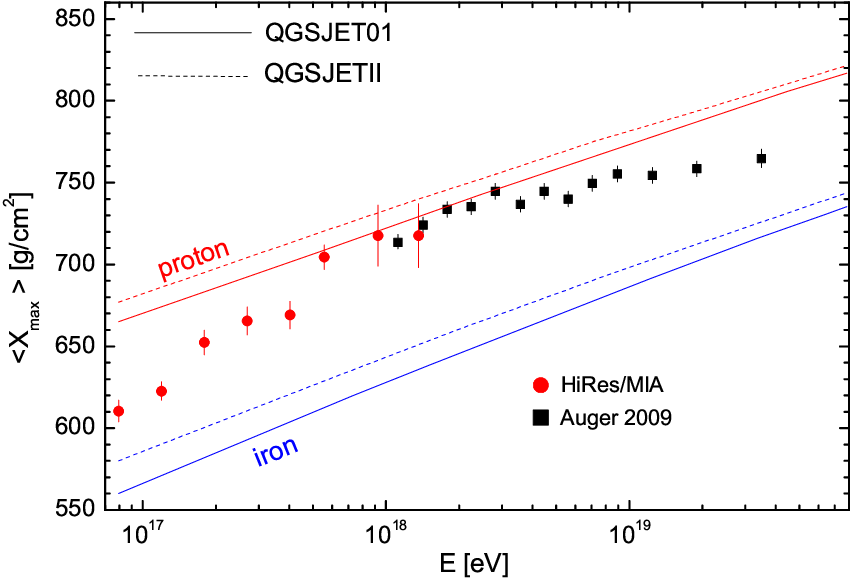}}\quad
\subfigure{\includegraphics[height=51mm]{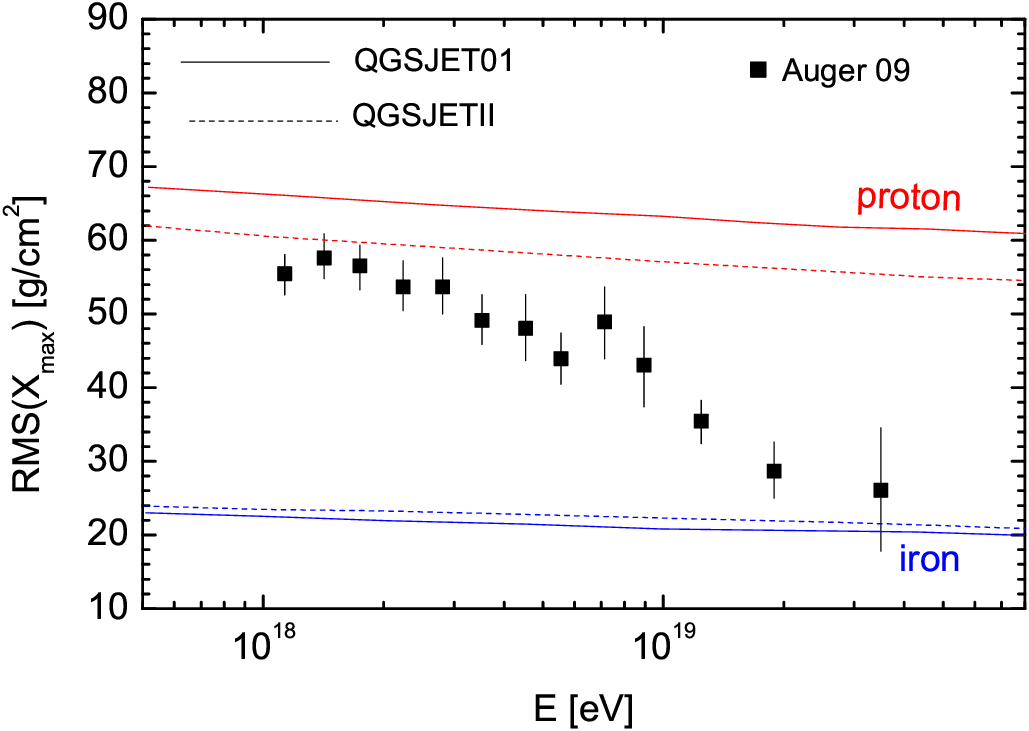}}}
\caption{
Auger data \cite{Auger-data,Abraham:2010yv,Bellido} on $X_{\rm max}$ as a 
function of energy (left panel) and on RMS($X_{\rm max}$), the width of 
distribution over $X_{\rm max}$, (right panel). The calculated 
values for protons and Iron are given according to QGSJETII model 
\cite{Ostapchenko:2005nj}. One can see from the right panel that 
RMS distribution becomes more narrow with energy increasing which 
implies the heavier composition.
} 
\label{Fig3}%
\end{figure*}

\begin{figure*}[ht]
\centering
\mbox{\subfigure{\includegraphics[height=50mm,width=65mm]{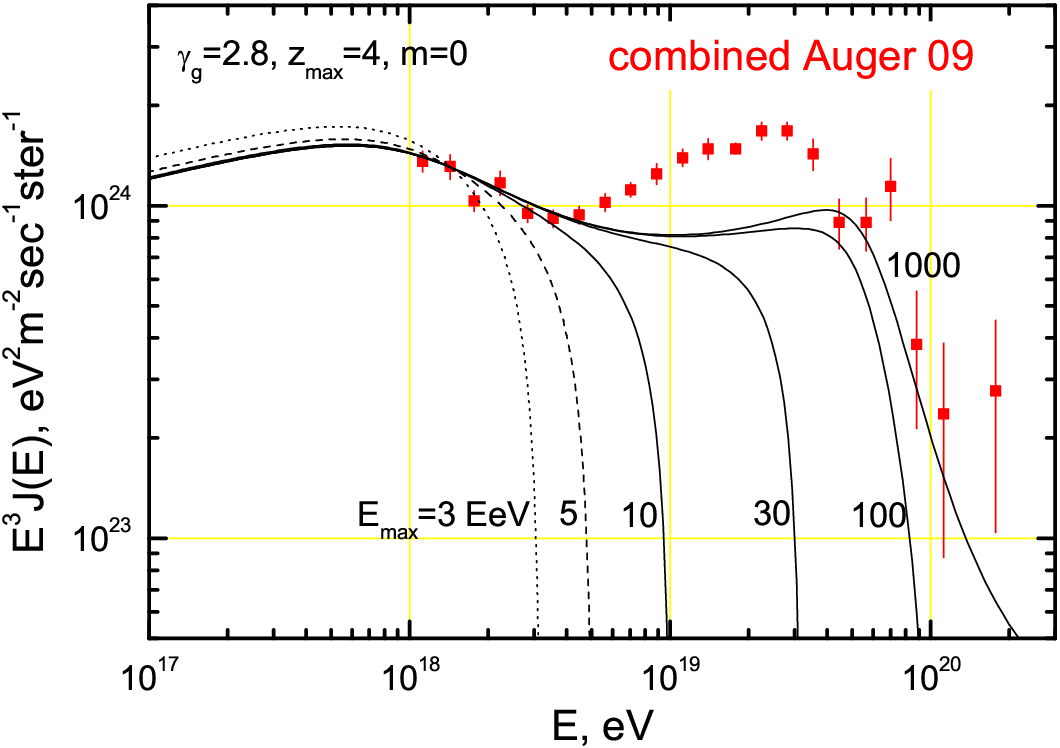}}\quad
\subfigure{\includegraphics[height=50mm,width=65mm]{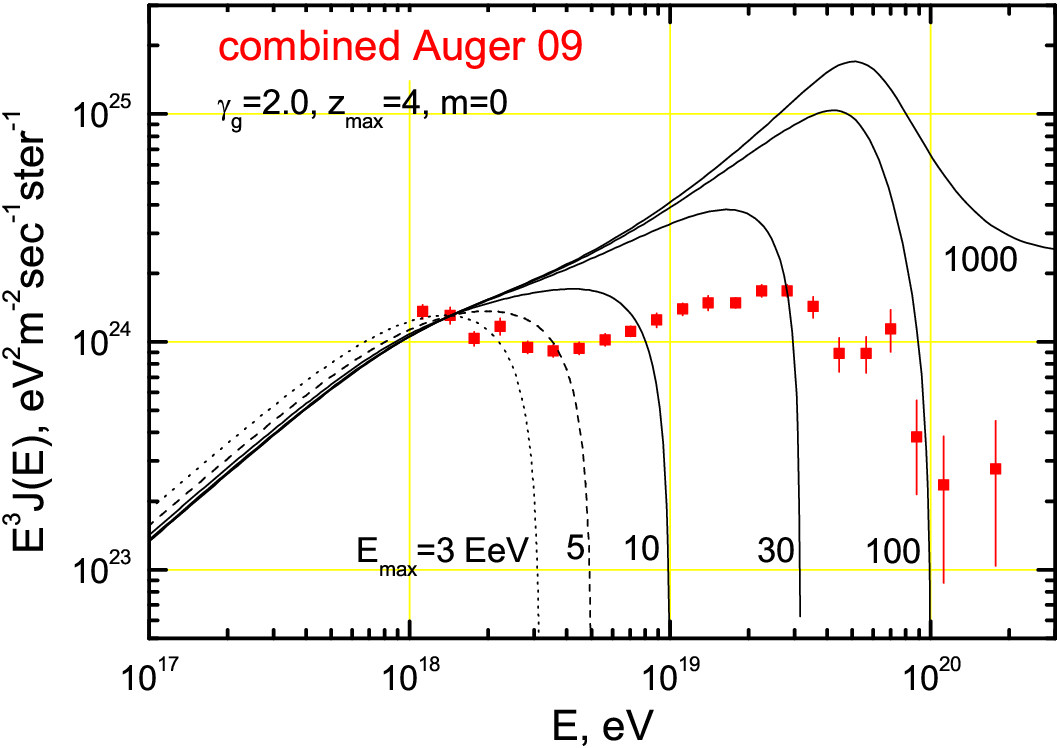}}}
\caption{
Comparison of calculated proton spectra with the combined Auger 
spectrum for different $E_p^{\rm max}$. Extreme cases $\gamma_g=2.8$ 
and 2.0 are shown in the left and right panels, respectively. The allowed 
$E_{\rm max}$ corresponds to curves lying below experimental points at 
$E \sim 4$~EeV, from which nuclei start dominating. 
} 
\label{Fig4}%
\end{figure*}

The Auger data on spectra and mass composition are quite different. In 
contrast to HiRes, the Auger data \cite{Auger-data,Abraham:2010yv,Bellido} 
show nuclei mass composition starting at 
$E \sim 4$~EeV, which becomes progressively heavier as energy 
increases (see Fig.~\ref{Fig3}). The change of the mass composition 
with energy is quite smooth, probably only with one peak at $E 
\approx 7$~EeV. It may indicate that the charge number $Z$ changes 
smoothly in sources. The energy spectrum has a sharp steepening at 
$E \sim (30 - 40)$~EeV (see squares in Fig.~\ref{Fig4}), but energy 
shape of this steepening, as our calculations show, is quite different 
from the one predicted for the GZK cutoff. This is not surprising 
taking into account the Auger mass composition. 

The width of the $X_{\rm max}$ distribution is a very powerful tool for 
determination of the UHECR mass composition \cite{Aloisio:2007rc}. It is free 
of many uncertainties involved in the $X_{\rm max}$ absolute value 
measuring method; the narrow width found in Auger experiment is 
difficult to falsify. On the other hand, the whole picture obtained 
by HiRes looks self-consistent. In particular, the confirmation of the 
pair-production dip in four experiments, including the early Auger data 
\cite{Berezinsky:2002nc,Aloisio:2006wv, Berezinsky:2002vt,Berezinsky:2005cq}, 
is a strong experimental argument in favor of the proton-dominated 
composition. The sources in this case can be AGN \cite{Berezinsky:2002vt} 
with a neutron mechanism of exit 
\cite{1977ICRC_Berezinsky,Mannheim:1998wp,Atoyan:2002gu}.

In this paper we concentrate on the consequences from the mass composition 
and energy spectrum measured {\em only} by the Auger detector and look 
for a natural model explaining these data. 

\section{Assumptions and features of the model}
\label{sec:model}
We consider a physical model directly following from the 
Auger observations with some additional ingredients. The accepted 
assumptions will be justified in the end of this section. 

The basic assumption in our model is the proton composition 
in the energy range ($1 - 3$)~EeV, which is supported by observations 
of both detectors Auger and HiRes (see Fig.~\ref{Fig3}, especially 
the left panel for the Auger data and Fig.~\ref{Fig2} for the HiRes data). 
We combine this observation with an additional assumption that these 
protons are extragalactic. The third ingredient of our model is an 
assumption of a rigidity-dependent acceleration in sources. 
In terms of maximum acceleration energy it may be formulated as 
$E_{\rm max}^{\rm acc}=ZE_0$, where $E_0$ is a universal energy to be
determined from data, and $Z$ is a nucleus charge number. These 
three assumptions complete the definition of the model. 

{\em The main approach of this work is as follows. 
We determine first the maximum acceleration energy for protons, 
$E_p^{\rm max}=E_0$. For this we calculate the extragalactic diffuse 
proton flux assuming the power-law generation spectrum 
$Q_g(E) \propto E^{-\gamma_g}$ with $E_{\rm max}=E_0$ and normalizing 
the calculated flux by the Auger flux at ($1 - 3$)~EeV. 
Varying $\gamma_g$ in the range $2.0 - 2.8$, we search for a maximum value 
of $E_0$ allowed by the Auger mass composition and energy spectrum.  
Increasing $E_0$ beyond this limit one comes to a contradiction 
either with mass composition or with energy spectrum in Fig.~\ref{Fig3}. 
}

The results are presented in Fig.~\ref{Fig4}. In calculations 
we use a homogeneous distribution of sources without cosmological 
evolution (evolution parameter $m=0$) and with maximal 
redshift $z_{\rm max}=4$. As a criterion of contradiction we choose 
an excess of calculated proton flux at energy ($4 - 5$)~EeV, where 
Auger data show the dominance of nuclei. The contradiction has 
different character for different values of $\gamma_g$. 

For steep source generation functions with $\gamma_g \simeq 
2.6 - 2.7$ the shape and flux of the Auger spectrum may be described 
by $E_p^{\rm max} \sim 10^{20} - 10^{21}$~eV; the contradiction occurs 
only in data on mass composition. The extreme case, given by 
$\gamma_g=2.8$, is displayed in the left panel of Fig.~\ref{Fig4}. 
In fact  all curves with $E_{\rm max} \geq 10$~EeV are below the 
data points at $E > 5$~EeV and hence compatible with Auger energy 
spectrum. However, these curves are excluded by prediction of the 
pure proton composition at $E \sim (4 - 5)$~EeV, i.e.\ due to 
contradiction in mass composition in a very narrow energy range 
(see  Fig.~\ref{Fig4}).

For flat generation spectra the contradiction is very pronounced: 
the predicted total proton flux exceeds the observed one. It can be 
seen in the right panel of Fig.~\ref{Fig4} for another extreme 
case, $\gamma_g=2.0$: for $E_{\rm max}= 5$~EeV the calculated proton 
flux exceeds the observed one at $E \approx 2$~EeV. 

We conclude with some redundancy that $E_p^{\rm max} \sim 
(4 - 10)$~EeV holds for all generation indices in the 
range 2.0 - 2.8. The maximum energy for Iron nuclei is  
$Z=26$ times higher and does not exceed 
$E_{\rm Fe}^{\rm max} \sim (100 - 300)$~EeV. 

In the end of this section we discuss our main assumption that 
the particles observed  at energy $(1 - 3)$~EeV are protons.  
Within experimental uncertainties they can be Helium. Does it exclude 
our conclusion about low $E_{\max}$ ?    

Our calculations with He nuclei show that, in contrast to protons,
we cannot fit well the three spectral points at energies $(1 - 3)$~EeV 
with good enough accuracy. The best, though not good, fit is given by 
$\gamma_g=2.8$ and $E^{\max}_{\rm He} = 4$~EeV. However, even this case has an
additional difficulty: the spectrum of Fe nuclei is too steep and
cannot explain the highest energy flux measured by Auger.  

\section{The Auger total energy spectrum}
\label{sec:tot}
\begin{figure*}[ht]
\centering
\mbox{\subfigure{\includegraphics[height=50mm,width=65mm]{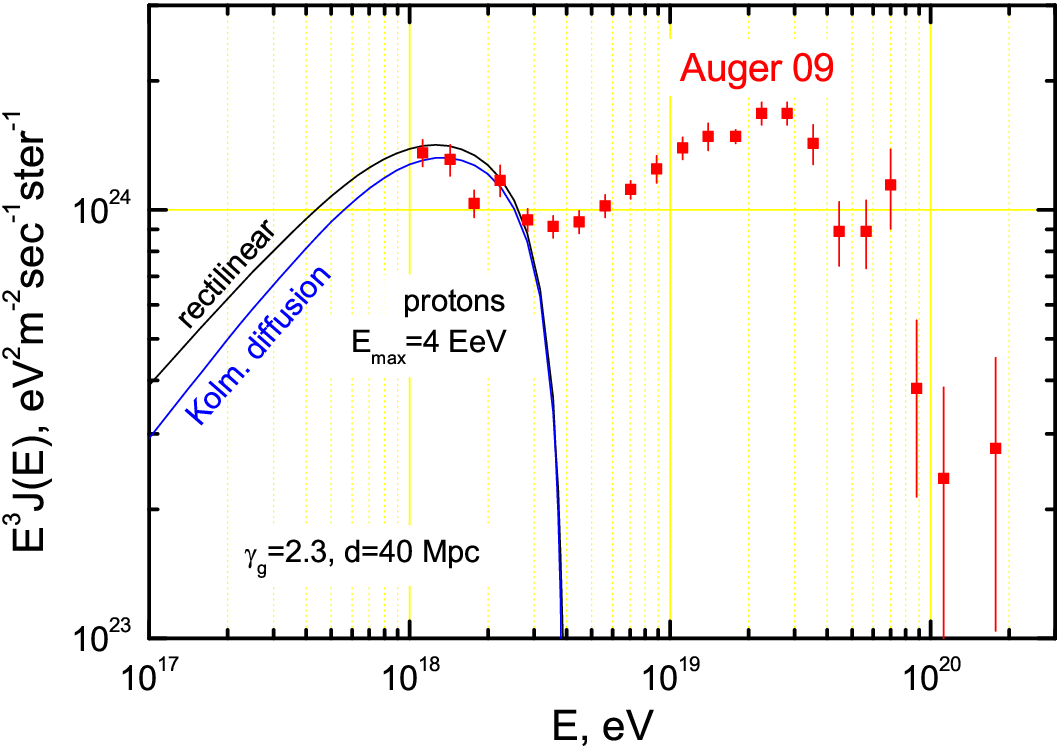}}\quad
\subfigure{\includegraphics[height=50mm,width=65mm]{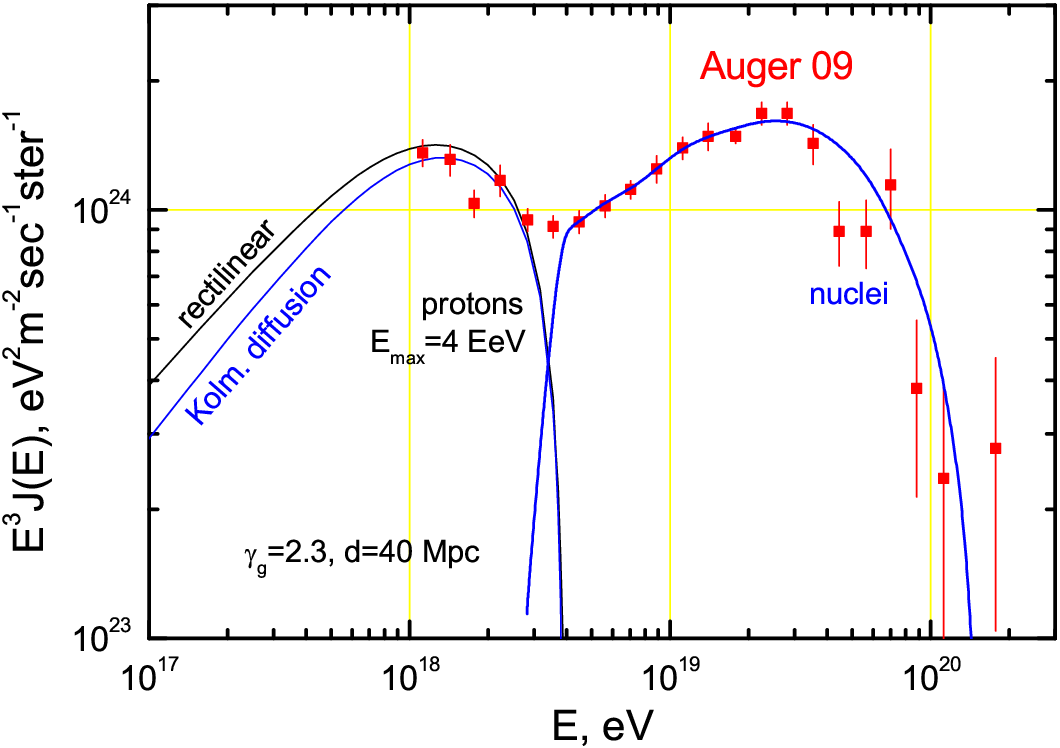}}}
\caption{
{\em Left panel:} Comparison of calculated proton spectra with the combined 
Auger spectrum for $\gamma_g=2.3$ and diffusive proton propagation (see 
text for details). The cutoff at $E_p^{\rm max} = 4$~EeV is needed to avoid 
the contradiction with data at $E> 3~$EeV. {\em Right panel:} Total UHECR 
spectrum in 'disappointing model' in comparison with the combined Auger 
spectrum. Spectrum of protons is taken from the left panel. The spectrum of 
nuclei is obtained by subtraction procedure as described in the text. 
} 
\label{Fig5}%
\end{figure*}

In Fig.~\ref{Fig5} we plot the calculated  total UHECR 
spectrum in the 'disappointing model', using $\gamma_g =2.3$, which 
might be the case for acceleration by relativistic shocks. The proton 
spectrum is calculated here in a diffusive model, the more realistic one for energies below 1~EeV. We assume the turbulent magnetic field with basic 
scales $(B_c,l_c)=(1 \mbox{ nG, } 1\mbox{ Mpc})$, the distance between sources 
$d \sim 40$~Mpc 
and the Kolmogorov diffusion coefficient (for notation and method of 
calculation see \cite{Berezinsky:2005fa,Aloisio:2008tx}). 
The analysis of proton maximum energy of acceleration (see left panel of 
Fig.~\ref{Fig5} ) gives $E_0=E_p^{\rm max}=4$~EeV, in a rough 
agreement with the analysis made for homogeneous distribution of sources. 
In fact this set of parameters gives the lowest $E_0$ allowed by Auger 
data, because a prediction for the maximum energy of Iron nuclei is 100~EeV, 
while in Auger one event with energy close to $200$~EeV was already observed. 
However, varying parameters, in particular increasing $\gamma_g$, it is easy 
to increase $E_0$ to ($5 - 6$)~EeV, and even 10~EeV cannot be excluded.  

The account for diffusion in extragalactic magnetic fields provides a 
flattening of the proton spectrum at $E \lesssim 1$~EeV, seen in 
Fig.~\ref{Fig5} as a 'diffusive cutoff', because flux $J(E)$ is multiplied 
by $E^3$. The 'diffusive cutoff' provides a transition from the steep galactic 
spectrum, most probably composed of Iron, to the flat spectrum of extragalactic 
protons. 

The spectrum of nuclei in Fig.~\ref{Fig5} is obtained by subtraction 
procedure first suggested in \cite{Berezinsky:2004wx}. We 
subtract the above-calculated proton spectrum from the total 
Auger spectrum. The resultant flux is plotted in the right panel of 
Fig.~\ref{Fig5} as a sum of different nuclei species. In the 
mixed composition model \cite{Allard:2005ha,Allard:2005cx} 
it looks quite possible  to fit the obtained nuclei spectrum by allowing for 
arbitrary fractions of primary nuclei in sources. 

The basic feature of the Auger mass composition, the
progressively heavier composition with energy increasing, is 
guaranteed in our model by the rigidity-dependent maximum energy of 
acceleration: at energy higher than $Z E_p^{\max}$ nuclei with charge 
$Z' < Z$ disappear, while heavier nuclei with larger $Z$ survive. Starting 
from $E_p^{\rm max} \sim (4 - 10)$~EeV, the higher energies are 
accessible only for nuclei with progressively larger values of $Z$. 
In particular, the maximum observed energy must correspond to Iron 
nuclei, which can reach $E \sim (100 - 300$)~EeV. In the next paper we plan 
to perform detailed calculations for diffusive propagation of nuclei with 
sources located in vertices of a cubic grid similar to 
\cite{Berezinsky:2007kz,Aloisio:2008tx}. 

\begin{figure*}[ht]
\centering
\mbox{\subfigure{\includegraphics[height=50mm,width=65mm]{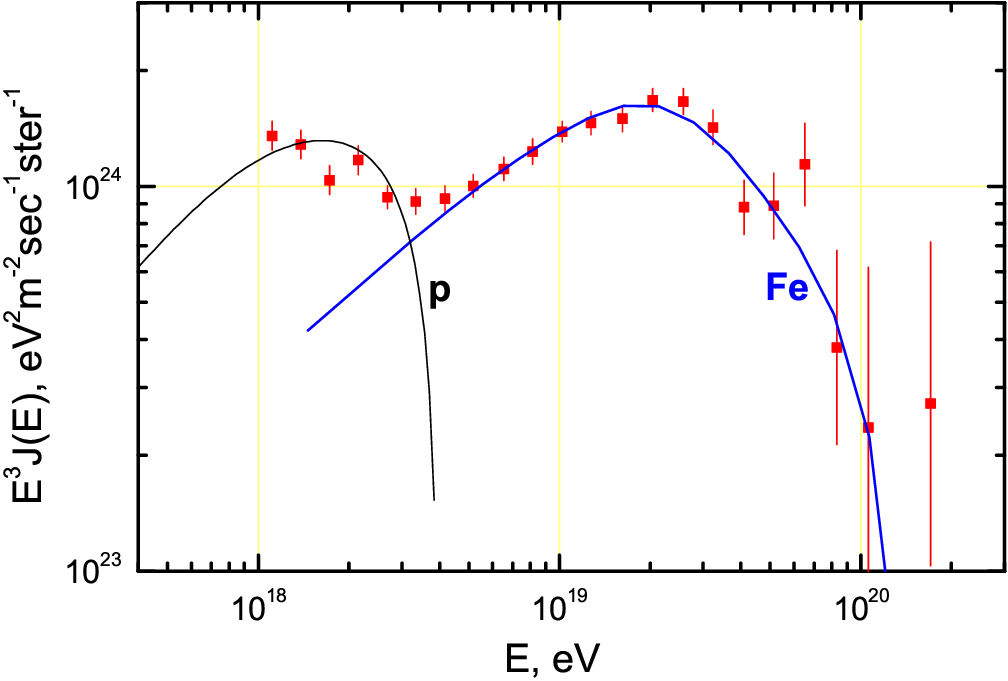}}\quad
\subfigure{\includegraphics[height=50mm,width=65mm]{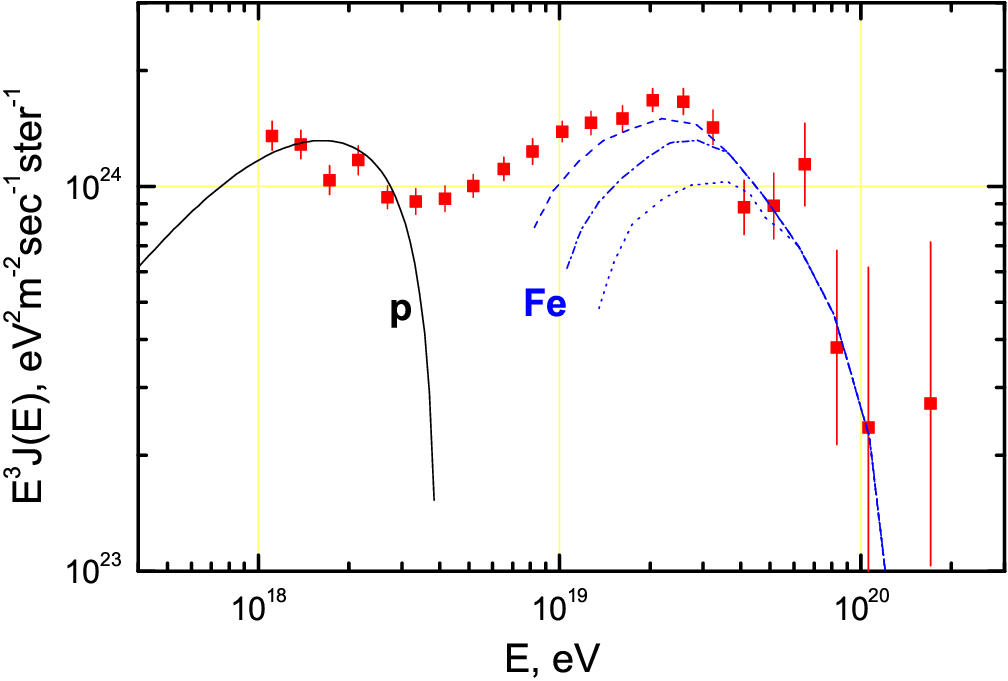}}}
\caption{
{\em Left panel:} The energy spectrum in two-component model with 
protons and Iron nuclei with $\gamma_g=2.0$ and $E_{\max}=4 Z$~EeV.
The Iron nuclei spectrum is calculated for homogeneous distribution of
the sources.
{\em Right panel:} As in the left panel, but with the `'diffusion cutoff' 
introduced for three different sets  of parameters $B_c, l_c, d$. The
gap between 2~EeV and $E_{\rm cut}$ (beginning of 'diffusive cutoff')
is expected to be filled by intermediate nuclei.   
} 
\label{Fig6}%
\end{figure*}

As a next step to exact calculations we consider a two-component model, 
with only protons and Iron nuclei being produced in sources. The
generation index $\gamma_g=2.0$ and the maximum acceleration energy 
$E_{\max}=4\,Z$~EeV. The proton spectrum is calculated as above and  
the primary Iron nuclei spectrum is calculated as in 
\cite{Aloisio:2008pp,Aloisio:2010he} for homogeneous distribution of 
sources. The spectra are presented in the left panel of Fig.~\ref{Fig6}. 
One may notice that the calculated spectrum of Iron describes well the 
cutoff in the Auger spectrum, which we failed to explain in many models 
of GZK cutoff. This steepening is caused by the photo-disintegration of 
Iron nuclei. The sharp acceleration cutoff at $E_{\rm Fe}^{\max} = 
ZE_p^{\max}$ is not shown in this figure.

To agree with Auger-observed mass composition, the Iron spectrum  
in Fig.~\ref{Fig6} must have a low-energy cutoff at $E \lesssim 
(20 - 30)$~EeV. Most naturally it is produced as a 'diffusive cutoff'  
which appears in models with lattice-located sources due to
{\em magnetic horizon}.

The magnetic horizon arises automatically in solutions of 
diffusion equations, e.g.\ in the Syrovatsky solution \cite{Aloisio:2004jda}, 
due to the Green function 
\begin{equation} 
G(E,\vec{r};E_g,\vec{r}_g) \propto \exp\left [-\frac{(\vec{r}-\vec{r}_g)^2}
{4\lambda (E,E_g)} \right ],
\label{green}
\end{equation}
where $\vec{r}$ and $\vec{r}_g$ are positions of an observer and a 
source, respectively, and $\lambda$ is given below by Eq.~(\ref{lambda}). 
Energies of primary Iron nuclei practically do not change during the lifetime of 
photo-disintegration $\tau$, and $\lambda$ can be given as 
\begin{equation}
\lambda(E,\tau)=\int_0^{\tau}dt D(E,t) \approx D(E)\,\tau, 
\label{lambda}
\end{equation}
where $D(E)$ is the diffusion coefficient at energy $E$. 
Thus, radius of the horizon and a condition when it cuts the spectrum off are  
given by 
\begin{equation}
r_{\rm hor}^2 = 4\, D(E)\, \tau \leq d^2,
\label{horizon}
\end{equation}
where $d$ is the lattice constant  (length), which is of order of distance
to the nearest source. 

For description of the diffusion we use as in \cite{Aloisio:2004jda} 
the magnetic field configuration $(B_c,l_c)= (1~{\rm nG}, 1~{\rm  Mpc})$, 
with the diffusion coefficient 
\begin{equation} 
D(E)=D_0 \left (\frac{E}{E_c}\right)^n,~~~~{\rm and}~~~~D_0=\frac{1}{3}\, c\, l_c,
\label{D}
\end{equation}
where $n=1/3$ at $E<E_c$, and $n=2$ at $E \gg E_c$, and 
\begin{equation}
E_c = Z\, e\, B_c\, l_c = 24 \times \frac{Z}{26}\times \frac{B_c}{1~{\rm nG}} 
\times \frac{l_c}{1~{\rm Mpc}}~{\rm EeV}.
\label{Ec}
\end{equation}
From Eqs.~(\ref{horizon}) - (\ref{D}) one obtains for the energy of cutoff 
$E_{\rm cut}$: 
\begin{equation}
\left (\frac{E_{\rm cut}}{E_c}\right )^n  =\frac{3d^2}{4c\tau l_c}= 
2.2 \times \left (\frac {d}{30~{\rm Mpc}} \right )^2 \times \left (\frac{10^9~{\rm yr}}
{\tau}\right ).
\label{Ecut/Ec}
\end{equation}
For $20 \lesssim d \lesssim 50$~Mpc and $\tau \sim (0.1 - 1)\times 10^9$~yr
\cite{Aloisio:2008pp,Aloisio:2010he}, we finally get 
\begin{equation}
E_{\rm cut} \sim E_c= 24 \times \frac{Z}{26} \times \frac{B_c}{\rm 1~nG} \times \frac{l_c}
{\rm 1~Mpc}~{\rm EeV}.
\label{Ecut}
\end{equation}

For the natural choice of parameters $B_c, l_c, d$ the beginning of
the 'diffusive cutoff' can start at $E_{\rm cut} \sim E_c$. 
When $E$ decreases, the $D(E)$ decreases too, and the horizon becomes 
less than distance to a nearby source. Still particles can arrive 
from there, being however stronger suppressed by the exponent in Eq.~(\ref{green}). 
Cutoff is expected to be sharp, but this statement expects a validation by 
numerical calculations.  

In the right panel of Fig. 6 we introduce in the Iron spectrum the low-energy 
'diffusive cutoff' for three different sets of parameters $B_c, l_c, d$.
The beginning of this cutoff $E_c$ for Iron nuclei is $Z=26$ times higher 
than for protons, i.e.\ $E_c \approx 2.6 \times 10^{19}$~eV, which has a 
reasonable physical meaning.  The gap between $2$~EeV and $26$~EeV is 
expected to be filled by intermediate nuclei. To provide a smooth RMS 
curve seen in the Auger data (Fig.~\ref{Fig3}) in this energy interval, 
we have many free parameters at our disposal in the form of arbitrary 
fractions of nuclei accelerated in a source. We plan to perform the 
detailed calculations in the future work.  

\section{Predictions and uncertainties}\label{signatures}
The predictions of our model are very disappointing for the future detectors. 

The maximum acceleration energy $E_{\rm max} \sim (100 - 300)$ EeV for Iron 
nuclei implies the energy per nucleon $E_p < E_{\rm max}/A \sim (2 - 5) \mbox{ EeV}$, 
well below the GZK cutoff for epochs with $z \lesssim 15$. Therefore, 
practically no cosmogenic neutrinos can be produced in collisions of 
protons and nuclei with CMBR photons. However neutrinos with $E_\nu \lesssim 
1\times 10^{17}$~eV can be generated in collisions of protons and nuclei with 
Extragalactic Background Light (EBL) photons; we will refer to these neutrinos 
as the EBL-produced ones. The main mechanism of their generation in our model 
is the decay of pions photo-produced by primary EeV protons on EBL. The 
EBL-produced neutrinos are also generated by nuclei after their 
photo-disintegration 
to nucleons. This mechanism provides a lower neutrino flux, because the 
secondary protons in our model are subdominant in comparison with the primary
ones. Finally, 
an additional neutrino flux appears at further lower energies due to decays of 
neutrons, the fragments of photo-disintegrated nuclei. The EBL-produced 
neutrino fluxes have been calculated under different assumptions in many
papers, see e.g.\ \cite{Engel:2001hd,Stanev:2007ry,Ave:2004uj,Hooper:2004jc,
Kotera:2010yn,Berezinsky:2010xa}.
 
For our case it was convenient to use the calculations of \cite{Kotera:2010yn}, 
where neutrino fluxes were given separately for production on CMBR, EBL and 
due to neutron decays. We used these calculations to estimate the upper limit 
for the EBL-produced neutrinos in our model. We found it to be 6 times below 
the upper limit which is expected to be obtained by IceCube after 5 years of 
observations \cite{Ahrens:2003ix}. The neutrino flux from neutron decays is 
less than this flux by two orders of magnitude.      

Correlation with UHECR sources also is absent due to deflection of nuclei 
in the galactic magnetic fields. The lack of correlation in the model is 
strengthened by dependence of the maximum energy on $Z$. 

The signatures of the 'disappointing model' for the Auger detector are the 
mass-energy relation, already seen in the elongation curve $X_{\rm max}(E)$, 
and transition from galactic to extragalactic cosmic rays  below the 
characteristic energy $E_c \sim 1$~EeV. 

There are some uncertainties in the model presented above. 
The most important one relates to estimates of $E_p^{\max}$. It is determined 
by the lowest energy where Auger data are inconsistent with proton composition 
(the 6th low-energy bin of the Auger data in Fig.~\ref{Fig3}). If this energy 
increases, $E_p^{\max}$ increases, too. The model collapses when allowed 
$E_p^{\max}$ reaches  e.g.\ ($50 - 100$)~EeV. Another case is given by the mass 
composition being heavy starting from 1~EeV. The cosmological evolution of sources 
are not included in our calculations. Since this effect slightly decreases 
$E_p^{\rm max}$, it is not needed to be taken into account. It is also possible 
that the EeV protons detected by Auger are secondary ones, i.e.\ those produced 
in photo-dissociation of primary nuclei in collisions with CMBR and extragalactic 
IR/UV photons. However, it was demonstrated in \cite{Aloisio:2008pp,Aloisio:2010he} 
that the flux of secondary protons in the EeV range is always smaller than the 
flux of parent primary nuclei. According to \cite{Allard:2008gj} it is considerably 
smaller than the sum of primary and secondary nuclei fluxes. 

\section{Discussions and conclusions}
\label{sec:conclusions}%
The suggested model is aimed at explanation of the observational data 
of the Auger detector only. The crucial for the model feature is (i) 
the proton composition in the energy range ($1 - 3$)~EeV, considered
in our paper as an assumption. Two additional assumptions are (ii) the 
extragalactic origin of the observed protons and (iii) their acceleration 
by the rigidity dependent mechanism with $E_{\rm max} = Z E_0$, where 
universal energy $E_0$ is the same for all nuclei. The upper limit 
on $E_0$ (maximum acceleration energy for protons) is obtained by 
calculating the proton spectrum at higher energies using the 
generation index $\gamma_g$ and normalizing flux at ($1 - 3$)~EeV 
by the Auger data. The cutoff of the proton spectrum at $E_0$ provides 
an agreement between the Auger flux and mass composition at 
higher energies. The calculations are performed for homogeneously 
distributed UHECR sources with cosmological evolution being neglected. 
Maximum redshift of sources is $z_{\rm max}=4$; indices of source 
power-law generation functions $\gamma_g$ vary in the wide range from 
$2.0$ to $2.8$. The obtained upper limit on $E_0$ is ($4 - 10$)~EeV. 
The maximum predicted energy corresponds to Iron and equals to 
($100 - 300$)~EeV. The maximum energy per nucleon is only ($2 - 5$)~EeV, 
and pion photo-production processes on CMBR are practically absent. 
Therefore, the GZK cutoff in the UHECR spectrum and the production 
of cosmogenic neutrinos are absent, too. The observed cutoff of 
the spectrum is provided by the nuclei photo-disintegration and is 
strengthened by the acceleration cutoff. 

The rigidity-dependent $E_{\rm max}$ provides the energy-de\-pendent mass 
composition: at energy higher than $ZE_0$ nuclei with charge $Z$ 
disappear, while heavier ones, with larger $Z$, survive. It agrees 
qualitatively with the Auger observations. This feature disfavors the 
correlation with UHECR sources at highest energies: at $100$~EeV Iron 
nuclei dominate in the spectrum and their deflection in the galactic 
magnetic fields prevents the correlation with possible sources. 

There are two signatures of this model for Auger data. The realistic 
one is the transition from galactic to extragalactic cosmic 
rays. This transition occurs due to intersection of the flat 
extragalactic proton spectrum with the steep one of galactic Iron 
nuclei. The flat spectrum of protons below 1~EeV appears due to 
diffusion of protons in extragalactic magnetic fields 
\cite{Lemoine:2004uw,Aloisio:2004fz}. The transition below 1~EeV 
combined with the nuclei composition above $(3 - 5)$~EeV is a strong 
signature of the model under discussion. 

The second signature is the energy-dependent mass composition, getting 
progressively heavier with energy increasing. The indication to this 
feature is already seen in the Auger data. 

Is there any alternative explanation of the EeV protons in the Auger data? 

The EeV protons could have galactic origin, though it contradicts 
the Standard Model for the galactic cosmic rays (see e.g.\ \cite{BeVo:2007}), 
where maximum acceleration energy, $E_{\rm max}^{\rm acc} \approx 
1\times 10^{17}$~eV, is attained by Iron nuclei. However, one may 
assume an additional high-energy proton component of the galactic cosmic 
rays extending up to ($2 - 3$)~EeV. The usual argument with anisotropy 
could be bypassed considering a case suggested in \cite{Wick:2003ex}. 
GRB could occur in our Galaxy $10^6 - 10^7$~years ago producing high 
energy protons by inner-shocks acceleration. The protons propagate in the 
mode of non-stationary diffusion, so that most of particles have already 
escaped from Galaxy and now only the tail of retarded particles 
with a reduced anisotropy is observed. In this model the transition 
from galactic to extragalactic cosmic rays occurs at ankle, which is 
clearly seen in the Auger data. This case differs from the ``standard'' 
ankle model, where transition occurs from galactic Iron to 
extragalactic protons; it needs a more detailed investigation. 

Another similar example is given by the model \cite{Calvez:2010uh}, where
all observed cosmic rays have galactic origin, and the Galactic GRBs 
are considered as sources. The primaries are assumed to be protons and Iron
nuclei, which dominate at low and highest energies, respectively. 

In conclusion, we face at present the most serious disagreement in the
observational data of the two biggest experiments in UHECR. 

HiRes observes signatures of proton propagation through CMBR in the form of 
the pair-production dip and GZK cutoff, and these observations are well 
confirmed by direct measurements of the proton-dominated mass 
composition. These observations predict a detectable accompanying UHE 
neutrino flux. The neutrino flux predicted in proton-dominated models 
\cite{Berezinsky:2010xa} may be registered by the future JEM-EUSO 
\cite{Takahashi:2009zzc} in the case of large $E_{\max}$ and strong cosmological
evolution of the sources. This detector can also observe the nearby UHECR 
sources using protons with energies up to $100$~EeV and above. 

Auger clearly observes the high-energy steepening of the spectrum, but 
its position and shape are rather different from the prediction of the 
GZK cutoff. The mass composition at $E \gtrsim 4$~EeV shows the dominance 
of nuclei becoming progressively heavier with energy increasing and 
probably reaching the pure Iron composition at $E \approx 35$~EeV. 
The Auger data allow the most conservative explanation in terms of low 
maximum-energy of acceleration, based on the observation of the EeV protons. 
It seriously ameliorates the problem of acceleration in astrophysical 
sources, diminishing the maximum energy down to $(100 - 300)$~EeV for 
Iron nuclei. This conservative and disappointing scenario can be 
confirmed by transition from galactic to extragalactic cosmic rays 
below $1$~EeV, by absence of cosmogenic neutrinos 
and by agreement with elongation rate calculated in this model. 

\section{Acknowledgments}
This work was presented at JEM-EUSO meeting on July 19-20 2009 and we are 
grateful to Alan Watson for encouraging us to write a paper and for 
valuable advices. We are very thankful to all participants of the 
SOCoR Workshop in Trondheim on June 15-18 2009, where the new Auger and 
HiRes data were presented and discussed, and a real brain attack on 
the problem was undertaken. We are grateful to Michael Kachelrie\ss, 
the main organizer of this remarkable workshop. We are also grateful to 
anonymous Referee for useful remarks. The work of A.Gazizov was 
supported in part by contract with Gran Sasso Center for Astroparticle 
Physics (CFA) funded by European Union and Regione Abruzzo under the 
contract P.O. FSE Abruzzo 2007-2013, Ob.\ CRO.


\begin{thebibliography}{10}
\expandafter\ifx\csname url\endcsname\relax
  \def\url#1{\texttt{#1}}\fi
\expandafter\ifx\csname urlprefix\endcsname\relax\def\urlprefix{URL }\fi
\expandafter\ifx\csname href\endcsname\relax
  \def\href#1#2{#2} \def\path#1{#1}\fi

\bibitem{Abbasi:2007sv}
R.~U. Abbasi, et~al., {Observation of the GZK cutoff by the HiRes experiment},
  Phys.\ Rev.\ Lett. 100 (2008) 101101.
\newblock \href {http://arxiv.org/abs/astro-ph/0703099}
  {\path{arXiv:astro-ph/0703099}}, \href
  {http://dx.doi.org/10.1103/PhysRevLett.100.101101}
  {\path{doi:10.1103/PhysRevLett.100.101101}}.

\bibitem{Abraham:2008ru}
J.~Abraham, et~al., {Observation of the suppression of the flux of cosmic rays
  above $4\times 10^{19}$eV}, Phys.\ Rev.\ Lett. 101 (2008) 061101.
\newblock \href {http://arxiv.org/abs/0806.4302} {\path{arXiv:0806.4302}},
  \href {http://dx.doi.org/10.1103/PhysRevLett.101.061101}
  {\path{doi:10.1103/PhysRevLett.101.061101}}.

\bibitem{Greisen:1966jv}
K.~Greisen, {End to the cosmic ray spectrum?}, Phys.\ Rev.\ Lett. 16 (1966)
  748--750.
\newblock \href {http://dx.doi.org/10.1103/PhysRevLett.16.748}
  {\path{doi:10.1103/PhysRevLett.16.748}}.

\bibitem{Zatsepin:1966jv}
G.~T. Zatsepin, V.~A. Kuzmin, {Upper limit of the spectrum of cosmic rays},
  JETP Lett. 4 (1966) 78--80.

\bibitem{Berezinsky:1988wi}
V.~S. Berezinsky, S.~I. Grigor'eva, {A Bump in the ultrahigh-energy cosmic ray
  spectrum}, Astron.\ Astrophys. 199 (1988) 1--12.

\bibitem{Berezinsky:2002nc}
V.~Berezinsky, A.~Z. Gazizov, S.~I. Grigorieva, {On astrophysical solution to
  ultra high energy cosmic rays}, Phys.\ Rev. D74 (2006) 043005.
\newblock \href {http://arxiv.org/abs/hep-ph/0204357}
  {\path{arXiv:hep-ph/0204357}}, \href
  {http://dx.doi.org/10.1103/PhysRevD.74.043005}
  {\path{doi:10.1103/PhysRevD.74.043005}}.

\bibitem{Aloisio:2006wv}
R.~Aloisio, et~al., {A dip in the UHECR spectrum and the transition from
  galactic to extragalactic cosmic rays}, Astropart.\ Phys. 27 (2007) 76--91.
\newblock \href {http://arxiv.org/abs/astro-ph/0608219}
  {\path{arXiv:astro-ph/0608219}}, \href
  {http://dx.doi.org/10.1016/j.astropartphys.2006.09.004}
  {\path{doi:10.1016/j.astropartphys.2006.09.004}}.

\bibitem{Berezinsky:2002vt}
V.~Berezinsky, A.~Z. Gazizov, S.~I. Grigorieva, {Signatures of AGN model for
  UHECR, }\href {http://arxiv.org/abs/astro-ph/0210095}
  {\path{arXiv:astro-ph/0210095}} (2002).

\bibitem{Berezinsky:2005cq}
V.~Berezinsky, A.~Z. Gazizov, S.~I. Grigorieva, {Dip in UHECR spectrum as
  signature of proton interaction with CMB}, Phys.\ Lett. B612 (2005) 147--153.
\newblock \href {http://arxiv.org/abs/astro-ph/0502550}
  {\path{arXiv:astro-ph/0502550}}, \href
  {http://dx.doi.org/10.1016/j.physletb.2005.02.058}
  {\path{doi:10.1016/j.physletb.2005.02.058}}.

\bibitem{sokol-trondheim}
P.~Sokolsky, {Results from HiRes}, in: {Searching for the Origins of Cosmic
  Rays (SOCoR 2009), http://web.phys.ntnu.no/\~{}mika/programme.html},
  Department of Physics, NTNU Trondheim, Norway, 2009.

\bibitem{Sokolsky:2010kb}
P.~Sokolsky, {Final Results from the High Resolution Fly's Eye (HiRes)
  Experiment, }\href {http://arxiv.org/abs/1010.2690} {\path{arXiv:1010.2690}}.

\bibitem{TA}
D.~Ikeda, Results from the telescope array experiment (invited), in: 22nd
  European Cosmic Ray Symposium,
  \href{http://ecrs2010.utu.fi/presentations_oral.shtml}{ECRS 2010}, University
  of Turku, Finland, Faculty of Education, University of Turku, 2010.

\bibitem{Thomson:2010tc}
G.~B. Thomson, {Results from the Telescope Array Experiment, }\href
  {http://arxiv.org/abs/1010.5528} {\path{arXiv:1010.5528}}.

\bibitem{Ostapchenko:2005nj}
S.~Ostapchenko, {Non-linear screening effects in high energy hadronic
  interactions}, Phys.\ Rev. D74 (2006) 014026.
\newblock \href {http://arxiv.org/abs/hep-ph/0505259}
  {\path{arXiv:hep-ph/0505259}}, \href
  {http://dx.doi.org/10.1103/PhysRevD.74.014026}
  {\path{doi:10.1103/PhysRevD.74.014026}}.

\bibitem{Sokolsky:2009Blois}
P.~Sokolsky, {New results from HIRES and Telescope Arrays}, in: {The XXIst
  Rencontres de Blois: Windows on the Universe,
  http://blois.in2p3.fr/2009/plenary\_sessions.html, June 2009}.

\bibitem{Abbasi:2009nf}
R.~U. Abbasi, et~al., {Indications of Proton-Dominated Cosmic Ray Composition
  above 1.6 EeV}, Phys.\ Rev.\ Lett. 104 (2010) 161101.
\newblock \href {http://arxiv.org/abs/0910.4184} {\path{arXiv:0910.4184}},
  \href {http://dx.doi.org/10.1103/PhysRevLett.104.161101}
  {\path{doi:10.1103/PhysRevLett.104.161101}}.

\bibitem{Auger-data}
M.~Unger, {Study of the Cosmic Ray Composition with the PAO}, in: {Searching
  for the Origins of Cosmic Rays,
  http://web.phys.ntnu.no/\~{}mika/programme.html, SOCoR 2009}, {Department of
  Physics, NTNU Trondheim, Norway}, 2009.

\bibitem{Abraham:2010yv}
J.~Abraham, et~al., {Measurement of the Depth of Maximum of Extensive Air
  Showers above $10^{18}$ eV}, Phys.\ Rev.\ Lett. 104 (2010) 091101.
\newblock \href {http://arxiv.org/abs/1002.0699} {\path{arXiv:1002.0699}},
  \href {http://dx.doi.org/10.1103/PhysRevLett.104.091101}
  {\path{doi:10.1103/PhysRevLett.104.091101}}.

\bibitem{Bellido}
J.~A. Bellido, {Measurement of the average depth of shower maximum and its
  fluctuations with the {Pierre Auger Observatory}}, in: {Proceedings of the
  \href{http://icrc2009.uni.lodz.pl/proc/html/index.php_id=7.html#HE.1.1} {31st
  ICRC, \L{}OD\.{Z} 2009}}, {University of \L{}od\.z with Andrzej Soltan
  Institute for Nuclear Studies}, 2009.

\bibitem{Aloisio:2007rc}
R.~Aloisio, V.~Berezinsky, P.~Blasi, S.~Ostapchenko, {Signatures of the
  transition from galactic to extragalactic cosmic rays}, Phys.\ Rev. D77
  (2008) 025007.
\newblock \href {http://arxiv.org/abs/0706.2834} {\path{arXiv:0706.2834}},
  \href {http://dx.doi.org/10.1103/PhysRevD.77.025007}
  {\path{doi:10.1103/PhysRevD.77.025007}}.

\bibitem{1977ICRC_Berezinsky}
V.~Berezinsky, {The origin of ultra high energy cosmic rays}, in: invited talk
  at 15th International Cosmic Ray Conference, Plovdiv, Bulgaria, Vol.~10,
  1977, pp. 84--107.

\bibitem{Mannheim:1998wp}
K.~Mannheim, R.~J. Protheroe, J.~P. Rachen, {On the cosmic ray bound for models
  of extragalactic neutrino production}, Phys.\ Rev. D63 (2001) 023003.
\newblock \href {http://arxiv.org/abs/astro-ph/9812398}
  {\path{arXiv:astro-ph/9812398}}, \href
  {http://dx.doi.org/10.1103/PhysRevD.63.023003}
  {\path{doi:10.1103/PhysRevD.63.023003}}.

\bibitem{Atoyan:2002gu}
A.~M. Atoyan, C.~D. Dermer, {Neutral beams from blazar jets}, Astrophys.\ J.
  586 (2003) 79--96.
\newblock \href {http://arxiv.org/abs/astro-ph/0209231}
  {\path{arXiv:astro-ph/0209231}}, \href {http://dx.doi.org/10.1086/346261}
  {\path{doi:10.1086/346261}}.

\bibitem{Berezinsky:2005fa}
V.~Berezinsky, A.~Z. Gazizov, {Diffusion of Cosmic Rays in Expanding Universe.
  (I)}, Astrophys.\ J. 643 (2006) 8--13.
\newblock \href {http://arxiv.org/abs/astro-ph/0512090}
  {\path{arXiv:astro-ph/0512090}}, \href {http://dx.doi.org/10.1086/502626}
  {\path{doi:10.1086/502626}}.

\bibitem{Aloisio:2008tx}
R.~Aloisio, V.~Berezinsky, A.~Gazizov, {Superluminal problem in diffusion of
  relativistic particles and its phenomenological solution}, Astrophys.\ J. 693
  (2009) 1275--1282.
\newblock \href {http://arxiv.org/abs/0805.1867} {\path{arXiv:0805.1867}},
  \href {http://dx.doi.org/10.1088/0004-637X/693/2/1275}
  {\path{doi:10.1088/0004-637X/693/2/1275}}.

\bibitem{Berezinsky:2004wx}
V.~S. Berezinsky, S.~I. Grigorieva, B.~I. Hnatyk, {Extragalactic UHE proton
  spectrum and prediction for iron-nuclei flux at 10**8-GeV to 10**9-GeV},
  Astropart.\ Phys. 21 (2004) 617--625.
\newblock \href {http://arxiv.org/abs/astro-ph/0403477}
  {\path{arXiv:astro-ph/0403477}}, \href
  {http://dx.doi.org/10.1016/j.astropartphys.2004.06.004}
  {\path{doi:10.1016/j.astropartphys.2004.06.004}}.

\bibitem{Allard:2005ha}
D.~Allard, E.~Parizot, E.~Khan, S.~Goriely, A.~V. Olinto, {UHE nuclei
  propagation and the interpretation of the ankle in the cosmic-ray spectrum},
  Astron.\ Astrophys. 443 (2005) L29--L32.
\newblock \href {http://arxiv.org/abs/astro-ph/0505566}
  {\path{arXiv:astro-ph/0505566}}.

\bibitem{Allard:2005cx}
D.~Allard, E.~Parizot, A.~V. Olinto, {On the transition from Galactic to
  extragalactic cosmic-rays: spectral and composition features from two
  opposite scenarios}, Astropart.\ Phys. 27 (2007) 61--75.
\newblock \href {http://arxiv.org/abs/astro-ph/0512345}
  {\path{arXiv:astro-ph/0512345}}, \href
  {http://dx.doi.org/10.1016/j.astropartphys.2006.09.006}
  {\path{doi:10.1016/j.astropartphys.2006.09.006}}.

\bibitem{Berezinsky:2007kz}
V.~Berezinsky, A.~Z. Gazizov, {Diffusion of cosmic rays in the expanding
  universe. II: Energy spectra of ultra-high energy cosmic rays}, Astrophys.\
  J. 669 (2007) 684--691.
\newblock \href {http://arxiv.org/abs/astro-ph/0702102}
  {\path{arXiv:astro-ph/0702102}}, \href {http://dx.doi.org/10.1086/520498}
  {\path{doi:10.1086/520498}}.

\bibitem{Aloisio:2008pp}
R.~Aloisio, V.~Berezinsky, S.~Grigorieva, {Analytic calculations of the spectra
  of ultra-high energy cosmic ray nuclei. I. The case of CMB radiation, }\href
  {http://arxiv.org/abs/0802.4452} {\path{arXiv:0802.4452}}.

\bibitem{Aloisio:2010he}
R.~Aloisio, V.~Berezinsky, S.~Grigorieva, {Analytic calculations of the spectra
  of ultra high energy cosmic ray nuclei. II. The general case of background
  radiation, }\href {http://arxiv.org/abs/1006.2484} {\path{arXiv:1006.2484}}.

\bibitem{Aloisio:2004jda}
R.~Aloisio, V.~Berezinsky, {Diffusive propagation of UHECR and the propagation
  theorem}, Astrophys.\ J. 612 (2004) 900--913.
\newblock \href {http://arxiv.org/abs/astro-ph/0403095}
  {\path{arXiv:astro-ph/0403095}}, \href {http://dx.doi.org/10.1086/421869}
  {\path{doi:10.1086/421869}}.

\bibitem{Engel:2001hd}
R.~Engel, D.~Seckel, T.~Stanev, {Neutrinos from propagation of ultra-high
  energy protons}, Phys.\ Rev. D64 (2001) 093010.
\newblock \href {http://arxiv.org/abs/astro-ph/0101216}
  {\path{arXiv:astro-ph/0101216}}, \href
  {http://dx.doi.org/10.1103/PhysRevD.64.093010}
  {\path{doi:10.1103/PhysRevD.64.093010}}.

\bibitem{Stanev:2007ry}
T.~Stanev, {Ultrahigh Energy Cosmic Rays and Neutrinos}, Nucl.\ Instrum.\ Meth.
  A588 (2008) 215--220.
\newblock \href {http://arxiv.org/abs/0711.1872} {\path{arXiv:0711.1872}},
  \href {http://dx.doi.org/10.1016/j.nima.2008.01.043}
  {\path{doi:10.1016/j.nima.2008.01.043}}.

\bibitem{Ave:2004uj}
M.~Ave, N.~Busca, A.~V. Olinto, A.~A. Watson, T.~Yamamoto, {Cosmogenic
  neutrinos from ultra-high energy nuclei}, Astropart.\ Phys. 23 (2005) 19--29.
\newblock \href {http://arxiv.org/abs/astro-ph/0409316}
  {\path{arXiv:astro-ph/0409316}}, \href
  {http://dx.doi.org/10.1016/j.astropartphys.2004.11.001}
  {\path{doi:10.1016/j.astropartphys.2004.11.001}}.

\bibitem{Hooper:2004jc}
D.~Hooper, A.~Taylor, S.~Sarkar, {The impact of heavy nuclei on the cosmogenic
  neutrino flux}, Astropart.\ Phys. 23 (2005) 11--17.
\newblock \href {http://arxiv.org/abs/astro-ph/0407618}
  {\path{arXiv:astro-ph/0407618}}, \href
  {http://dx.doi.org/10.1016/j.astropartphys.2004.11.002}
  {\path{doi:10.1016/j.astropartphys.2004.11.002}}.

\bibitem{Kotera:2010yn}
K.~Kotera, D.~Allard, A.~V. Olinto, {Cosmogenic Neutrinos: parameter space and
  detectability from PeV to ZeV}, JCAP 1010 (2010) 013.
\newblock \href {http://arxiv.org/abs/1009.1382} {\path{arXiv:1009.1382}},
  \href {http://dx.doi.org/10.1088/1475-7516/2010/10/013}
  {\path{doi:10.1088/1475-7516/2010/10/013}}.

\bibitem{Berezinsky:2010xa}
V.~Berezinsky, A.~Gazizov, M.~Kachelriess, S.~Ostapchenko, {Fermi-LAT
  restrictions on UHECRs and cosmogenic neutrinos, }\href
  {http://arxiv.org/abs/1003.1496} {\path{arXiv:1003.1496}}.

\bibitem{Ahrens:2003ix}
J.~Ahrens, et~al., {Sensitivity of the IceCube detector to astrophysical
  sources of high energy muon neutrinos}, Astropart.\ Phys. 20 (2004) 507--532.
\newblock \href {http://arxiv.org/abs/astro-ph/0305196}
  {\path{arXiv:astro-ph/0305196}}, \href
  {http://dx.doi.org/10.1016/j.astropartphys.2003.09.003}
  {\path{doi:10.1016/j.astropartphys.2003.09.003}}.

\bibitem{Allard:2008gj}
D.~Allard, N.~G. Busca, G.~Decerprit, A.~V. Olinto, E.~Parizot, {Implications
  of the cosmic ray spectrum for the mass composition at the highest energies},
  JCAP 0810 (2008) 033.
\newblock \href {http://arxiv.org/abs/0805.4779} {\path{arXiv:0805.4779}},
  \href {http://dx.doi.org/10.1088/1475-7516/2008/10/033}
  {\path{doi:10.1088/1475-7516/2008/10/033}}.

\bibitem{Lemoine:2004uw}
M.~Lemoine, {Extra-galactic magnetic fields and the second knee in the
  cosmic-ray spectrum}, Phys.\ Rev. D71 (2005) 083007.
\newblock \href {http://arxiv.org/abs/astro-ph/0411173}
  {\path{arXiv:astro-ph/0411173}}, \href
  {http://dx.doi.org/10.1103/PhysRevD.71.083007}
  {\path{doi:10.1103/PhysRevD.71.083007}}.

\bibitem{Aloisio:2004fz}
R.~Aloisio, V.~S. Berezinsky, {Anti-GZK effect in UHECR diffusive propagation},
  Astrophys.\ J. 625 (2005) 249--255.
\newblock \href {http://arxiv.org/abs/astro-ph/0412578}
  {\path{arXiv:astro-ph/0412578}}, \href {http://dx.doi.org/10.1086/429615}
  {\path{doi:10.1086/429615}}.

\bibitem{BeVo:2007}
E.~G. Berezhko, H.~J. {V\"olk}, {Spectrum of cosmic rays, produced in supernova
  remnants}, Astrophys.\ J. 661 (2007) L175--L177.
\newblock \href {http://arxiv.org/abs/0704.1715} {\path{arXiv:0704.1715}}.

\bibitem{Wick:2003ex}
S.~D. Wick, C.~D. Dermer, A.~Atoyan, {High-energy cosmic rays from gamma-ray
  bursts}, Astropart.\ Phys. 21 (2004) 125--148.
\newblock \href {http://arxiv.org/abs/astro-ph/0310667}
  {\path{arXiv:astro-ph/0310667}}, \href
  {http://dx.doi.org/10.1016/j.astropartphys.2003.12.008}
  {\path{doi:10.1016/j.astropartphys.2003.12.008}}.

\bibitem{Calvez:2010uh}
A.~Calvez, A.~Kusenko, S.~Nagataki, {Role of Galactic sources and magnetic
  fields in forming the observed energy-dependent composition of
  ultrahigh-energy cosmic rays}, Phys.\ Rev.\ Lett. 105 (2010) 091101.
\newblock \href {http://arxiv.org/abs/1004.2535} {\path{arXiv:1004.2535}},
  \href {http://dx.doi.org/10.1103/PhysRevLett.105.091101}
  {\path{doi:10.1103/PhysRevLett.105.091101}}.

\bibitem{Takahashi:2009zzc}
Y.~Takahashi, {The JEM-EUSO mission}, New J. Phys. 11 (2009) 065009.
\newblock \href {http://dx.doi.org/10.1088/1367-2630/11/6/065009}
  {\path{doi:10.1088/1367-2630/11/6/065009}}.

\end{thebibliography}

\end{document}